\documentclass[submission,copyright]{eptcs}
\providecommand{\event}{TFPIE} % Name of the event you are submitting to

\usepackage[numbers]{natbib}
\usepackage{xspace}
\usepackage{alltt}

\def\htdp{\emph{How to Design Programs}\xspace}
\def\htdc{\emph{How to Design Classes}\xspace}
\def\profj{{ProfessorJ}\xspace}

\newcommand\thetitle{From Principles to Practice with Class in the First Year}

\title{\thetitle}
\author{Sam Tobin-Hochstadt \quad\qquad David Van Horn
\institute{Northeastern University\\
Boston, Massachusetts, USA}
\email{\{samth,dvanhorn\}@ccs.neu.edu}
}
\def\titlerunning{\thetitle}
\def\authorrunning{S. Tobin-Hochstadt \& D. Van Horn}
\begin{document}
\maketitle

\begin{abstract}
We propose a bridge between functional and object-oriented programming
in the first-year curriculum.  Traditionally, curricula that begin
with functional programming transition to a professional, usually
object-oriented, language in the second course.  This transition poses
obstacles for students, and often results in confusing the details of
development environments, syntax, and libraries with the fundamentals
of OO programming that the course should focus on.  Instead, we
propose to begin the second course with a sequence of custom teaching
languages which minimize the transition from the first course, and
allow students to focus on core ideas.  After working through the
sequence of pedagogical languages, we then transition to Java, at
which point students have a strong command of the basic principles.
We have 3 years of experience with this course, with notable success.
\end{abstract}

\section{Introduction}
\label{sec:intro}

Many universities and colleges aim to teach their students proficiency
in an industrial object-oriented programming language by the end of
the students' first year.  The most common approach to achieve this
aim is to teach an industrial language in the first course, often
Java, starting on the first day.  Other curricula take a more indirect
route by teaching functional programming in the first semester,
followed by a second semester in Java.  The latter approach is an
improvement over the first, as pointed out by numerous
observers~\cite{dvanhorn:Felleisen2004Structure, local:spolsky,
  dvanhorn:Chakravarty2004Risks, dvanhorn:Ragde2008Chilling}, but
both suffer serious flaws.

As an example, Northeastern University teaches introductory programming
in the first semester using \emph{How to Design
  Programs}~\cite{dvanhorn:Felleisen2001How}, followed by
object-oriented programming using \emph{How to Design
  Classes}~\cite{local:htdc} in the second semester.  This sequence
was designed to provide a smooth path for incoming students with
 a competence in high-school level algebra to reach
proficiency in Java by the end of their first year
\cite{dvanhorn:Felleisen2004Structure}.  It was a major improvement
over the previous Java-first curriculum in terms of student success,
attrition, and preparation for subsequent
courses~\cite{dvanhorn:Proulx2006Design}.  However, significant
problems remain; in particular, the second semester course violates
the designers' own principles (as recalled
in~\cite{dvanhorn:Bloch2000Scheme}):
\begin{enumerate}
\item \emph{introduce only those language constructs that are necessary to
  teach programming principles}, and
\item \emph{choose a language with as
  few language constructs as possible, and one in which they can be
  introduced one at a time}.
\end{enumerate}

The problem is that the first semester ends with an advanced
pedagogical functional language and the second semester starts with
Java, although it focuses on a side-effect free subset.  Despite this focused
subset, this transition is too abrupt to meaningfully bridge the gap
between functional and object-oriented programming, because
several other significant transitions happen in concert:
\begin{itemize}
\item from a highly regular and minimal syntax to a complicated
  irregular syntax,

\item from an untyped language to a typed language,

\item from a pedagogical programming environment (DrRacket) to a professional
programming environment (Eclipse),

\item from a language with numeric values corresponding to
  mathematical objects to a language with numeric values corresponding
  to common machine representations,\footnote{While this may seem like
    a minor point, details of numeric representation can crop up
    quickly in a Java-based course---for example, $1/3$ cannot be
    represented by any Java numeric types.}

\item from a language with image literals and graphical libraries to
  one in which graphical programming is tedious,

\item from an interaction-oriented language and tool suite to a
  compiled, batch-oriented language.
\end{itemize}

This abrupt transition has several negative consequences which we have
experienced first-hand: the principles of object-oriented programming
are obscured and de-emphasized, struggling with the programming
environment is frustrating and can cause potentially good students to
leave the program, it favors students with prior exposure to the
particular tools (a set that is demographically skewed), it inhibits
students from experimenting by allowing them to rely upon past skills,
and it creates the false impression that courses are discrete units of
instruction that can be discarded after successful completion rather
than being part of a continuous and cumulative educational experience.

We contribute an alternative approach to the second semester that
overcomes these problems and provides a gradual introduction 
to object-oriented programming.  Our approach
starts the second semester by introducing \emph{only} the
concept of programming with objects, while all other aspects of the course
remain where they were left off in the previous semester.  This
 allows other concepts to be introduced at the point at
which they are relevant and motivated.  Despite this more gradual
approach, the course accomplishes the goal of reaching industrial
language competence by the end of the semester, covering a super-set
of the concepts and topics covered in the \emph{How to Design Classes}-based
course.

\paragraph{Outline} The remainder of this paper is organized as follows:
 section~\ref{sec:background} provides background on \emph{How to
  Design Programs} and the context and constraints involved in the
first year at Northeastern.  Section~\ref{sec:shift} describes our
approach to the second semester, which starts with a small shift in
perspective to bridge the gap between functional programming and
object oriented programming.  Section~\ref{sec:industrial} describes
the path from our pedagogical languages to an industrial object-oriented
programming language.  Section~\ref{sec:related-work} discusses the
relation to existing work and section~\ref{sec:conclusion} concludes.

%% \cite{local:dpc}

%% Some of these are premature, and better introduced later in the
%% course.  Some are just irrelevant trivia (different syntax) that often
%% favors ``experienced'' students while needlessly intimidating students
%% with only the previous semester under their belt.  Many of these
%% trivial matters, e.g. Eclipse configuration, javac command line
%% arguments, the CLASSPATH, can be very frustrating to the uninitiated,
%% causing students to believe struggling with such crap is \emph{the
%%   essence of computer science}.  Bright, promising students may leave
%% the program, understandably so, in order to pursue more intellectually
%% fulfilling subjects.

\section{Background: the context at Northeastern}
\label{sec:background}

At Northeastern, the College of Computer \& Information Science (CCIS)
requires a four course introductory sequence for Computer Science majors in the first year.  The
first semester features both a course on discrete mathematics and
 an introduction to programming following the \emph{How to Design
  Programs} curriculum.  The second semester follows with a course on
object-oriented programming and one featuring formal reasoning about
programs, both on paper and with the ACL2 theorem
prover~\cite{dvanhorn:Kaufmann2000ComputerAided}.

After the first year, students take a wide variety of
follow-up courses, ranging from a required course in ``Object-oriented
design'' to architecture, operating systems, robotics, and programming
languages.  No standard language is used in these courses.  

More significantly, Northeastern distinctively emphasizes experiential
education, with almost all Computer Science majors participating in a
6 month ``co-op'' internship after their third semester.  These co-ops
take place at a wide variety of companies, and while most students do
some software development, there is no single platform or
set of tools that commands majority use.
In particular, there is wide variation in the languages students use
while on co-op. This combination sets the constraints under which we
designed our approach.

\subsection{A first course on \emph{How to Design Programs}}

In the first semester, students are introduced to the ``design
recipe'', a step-by-step process for going from English descriptions
of problems to working programs.  The design recipe involves six steps:
\begin{enumerate}
\item Analyze the information involved in the problem and express
  how to represent it as data.
\item Write down a function signature, a summary of the purpose
  of the function, and a function stub.
\item Illustrate the signature and the purpose statement with some
  functional examples.
\item Take an inventory of the input data that can be used to compute
  an answer.
\item Write the code for the function.
\item Verify the behavior of the program against the functional
  examples given earlier.
\end{enumerate}

Students explore program design using this process in the context of a
series of successively richer pedagogical programming language
levels~\cite{dvanhorn:Felleisen2004Structure,
  dvanhorn:Felleisen2001How} that are included in the DrRacket
(formerly DrScheme) programming
environment~\cite{dvanhorn:Findler2002DrScheme}.
These languages are called the beginning student language (BSL), the
intermediate student language (ISL), and the advance student language
(ASL).
The language and environment include several tools in support of the
design recipe.  For example, functional examples can be written as
executable tests by writing \texttt{check-expect}
expression~\cite{local:check-expect}; an algebraic stepper and REPL
are provided to interact with programs at each step of the design
process.

Finally, the first semester course makes extensive use of a library
for developing interactive animations and games using functional
programming and functional
graphics~\cite{dvanhorn:Felleisen2009Functional, local:barland-sfp10}.

\subsection{The goal of the second course}

After the second course, students should both (a) be prepared for
subsequent courses in the curriculum, which expect familiarity with
Java and standard Java libraries, (b) be prepared for co-ops in which
they will use professional-grade languages and tools which will almost
certainly be object-oriented.  More significantly, we aim to teach the
key insights behind the object-oriented approach to program design.

These constraints, while in detail specific to Northeastern and the
CCIS curriculum, are broadly similar to the requirements for the first
year at many universities.  Our course also attends to smaller and
more idiosyncratic elements of our curriculum, ranging from formal
reasoning to algorithmic analysis, as described in the following
sections.

\section{A small shift of focus}
\label{sec:shift}

On the first day of the second semester, we introduce a single
linguistic concept to an otherwise unchanged context of the previous
semester: the idea of an object.
An object is a new kind of value that can, as a first cut, be
understood as a pairing together of two familiar concepts: data and
functionality.

\begin{itemize}
\item An object is like a structure in that it has a fixed number of
fields, thus an object (again, like a structure) can represent
compound data. But unlike a structure, an object contains not just
data, but functionality too;

\item An object is like a (set of) function(s) in that it has behavior---it computes; it is
not just inert data.
\end{itemize}

This suggests that objects are a natural fit for well-designed programs
since good programs are organized around data definitions and
functions that operate over such data. An object, in essence, packages
these two things together into a single programming apparatus. This
has two important consequences:

\begin{enumerate}

\item Students already know how to design programs oriented around objects.

Since objects are just the combination of two familiar concepts that
students already use to design programs, they already know how to
design programs around objects, even if they have never heard the term
``object'' before.

\item Objects enable new kinds of abstraction and composition.

Although the combination of data and functionality may seem simple,
objects enable new forms of abstraction and composition. That is,
objects open up new approaches to the construction of computations. By
studying these new approaches, we can distill new design
principles. Because we understand objects are just the combination of
data and functionality, we can understand how all of these principles
apply in the familiar context of programming with functions. 
\end{enumerate}

\subsection{The basics of objects}

To begin with, we introduce the notion of a \emph{class definition},
which can be thought of at first as a structure definition in that
it defines a new class of compound data.  A class is defined using the
{\tt define-class} form:
\begin{verbatim}
    (define-class posn (fields x y))
\end{verbatim}
This is similar to the {\tt define-struct} form of the first semester,
used as follows:
\begin{verbatim}
    (define-struct posn (x y))
\end{verbatim}

An \emph{object} is a value that is a member of this class of data,
which can be constructed with the {\tt new} keyword, a class name, and
the appropriate number of arguments for the fields of the object:
\begin{verbatim}
    (new posn 3 4)
\end{verbatim}
An object understands some set of \emph{messages}.  Simple
structure-like objects understand messages for accessing their fields
and messages are sent by using the {\tt send} keyword, followed by an object,
a message name, and some number of arguments:
\begin{verbatim}
    (send (new posn 3 4) x) ;=> 3
    (send (new posn 3 4) y) ;=> 4
\end{verbatim}
The {\tt send} notation is simple, but syntactically heavy.  Once
students are comfortable with {\tt send}, we introduce 
shorthand to make it more convenient, writing {\tt (x . m)} for {\tt
  (send x m)}.  The \emph{dot notation} can be nested, so {\tt (x . m
  . n)} is shorthand for {\tt (send (send x m) n)}.  (The approach of
introducing a simple, uniform syntax and later introducing a
convenient shorthand that would have been confusing to start with
follows the approach of first introducing {\tt cons} and then later {\tt list}
and {\tt quote} in the first semester.)

It is possible to endow objects with functionality by defining \emph{methods},
which extend the set of messages an object understands.  A method definition
follows the same syntax as a function definition, but is located inside of a
class definition.  Here is a more complete development of the {\tt posn} class
with two methods:
\begin{verbatim}
    ;; A Posn is a (new posn Number Number),
    ;; which represents a point on the Cartesian plane
    (define-class posn (fields x y)

      ;; dist : Posn -> Number
      ;; Distance between this posn and that posn
      (check-expect ((new posn 0 0) . dist (new posn 3 4)) 5)
      (define (dist that)
        (sqrt (+ (sqr (- (this . x) (that . x)))
                 (sqr (- (this . y) (that . y))))))

      ;; dist-origin : -> Number
      ;; Distance of this posn from the origin
      (check-expect ((new posn 0 0) . dist-origin) 0)
      (check-expect ((new posn 3 4) . dist-origin) 5)
      (define (dist-origin)
        (this . dist (new posn 0 0))))
\end{verbatim}

This class definition defines a new class of values which are {\tt
  posn} objects.  Such objects are comprised of two numeric values and
understand the messages {\tt x}, {\tt y}, {\tt dist}, and {\tt
  dist-origin}.  Unit tests have been included with each method
definition, following the principles of the design recipe studied in
the first semester.  Although {\tt check-expect} forms can appear
within class definitions, they are lifted to the top-level when a
program is run.

Methods can be defined to consume any number of arguments, but they
are implicitly parameterized over {\tt this}, the object that received
the message.

\subsection{Where did the {\tt cond} go?}

Unions, and recursive unions in particular, are a fundamental kind of
data definition that students are well-versed in from the previous
semester.  A fundamental early lesson is how to represent (recursive)
unions using classes and how to write recursive methods.  As an
example, figure~\ref{fig:tree} defines binary trees of numbers
(an archetypal recursive union data definition)
using the Beginning Student language (BSL) as used at the start of the
first semester, and also using the Class language of our course.
%% \begin{verbatim}
%%     ;; A list of numbers (LoN) is one of:
%%     ;; - (new empty%)
%%     ;; - (new cons% Number LoN)

%%     (define-class empty%
%%       ;; Length of this empty list
%%       (check-expect ((new empty%) . len) 0)
%%       (define (len) 0))

%%     (define-class cons% (fields first rest)
%%       ;; Length of this cons list
%%       (check-expect ((new cons% 7 (new empty%)) . len) 1)
%%       (define (len)
%%         (add1 (this . rest . len))))
%% \end{verbatim}

\begin{figure}[h!]
\begin{minipage}[t]{3.5in}
\begin{verbatim}
#lang bsl
;; A Tree is one of:
;; - (make-leaf Number)
;; - (make-node Tree Number Tree)
(define-struct leaf (v))
(define-struct node (left v right))

;; sum : Tree -> Number
;; sums the elements of the given tree
(define (sum a-tree)
  (cond [(leaf? a-tree) (leaf-v a-tree)]
        [else
         (+ (sum (node-left  a-tree))
            (node-v a-tree)
            (sum (node-right a-tree)))]))




(check-expect (sum (make-leaf 7)) 7)
(check-expect 
  (sum (make-node
         (make-leaf 1)
         5 
         (make-node (make-leaf 0)
                    10
                    (make-leaf 0))))
  16)
\end{verbatim}
\end{minipage}
\begin{minipage}[t]{3in}
\begin{verbatim}
#lang class/1
;; A Tree is one of:
;; - (new leaf Number)
;; - (new node Tree Number Tree)
;; and implements
;; sum : -> Number
;; sums the elements of this tree

(define-class leaf
  (fields v)
  (define (sum) (this . v)))

(define-class node
  (fields left v right)
  (define (sum)
    (+ (this . left . sum)
       (this . v)
       (this . right .sum))))

(check-expect ((new leaf 7) . sum) 7)
(check-expect 
  ((new node
        (new leaf 1)
        5 
        (new node (new leaf 0) 
                  10
                  (new leaf 0))))
    . sum)
  16)
\end{verbatim}
\end{minipage}
\caption{Binary tree sum in Beginning Student and in the Class language}
\label{fig:tree}
\end{figure}

The structure of this program is analogous to the approach of
the previous semester but this example brings to light an important
difference with the functional approach.  The method for computing the
sum of a leaf is defined in the {\tt leaf} class, while the
method for computing the sum of a node is in the {\tt node} class.
When a tree object is sent the {\tt sum} method, there is no function
with a conditional to determine whether the object is a
leaf---instead, the object itself takes care of computing the sum
based on its own {\tt sum} method.  This shift in perspective
is at the core of object-orientation: objects contain their own
behavior and the case analysis previously done in functions is
eliminated.

%% The semester starts with the minimal conceptual change from the
%% previous semester: the addition of objects (same IDE, same syntax,
%% same semantics [just a new kind of value], same interactive video
%% games, etc.).

\subsection{Worlds and animations}

At Northeastern, Programming in the first semester is often oriented
around interactive event-driven video games.  The basic design of a
video game involves defining a data representation for states of the
game and functions for transitioning between states based on events
such as clock ticks, keyboard input, or mouse events.  The design of a
game thus involves the design of data and operations on that data; in
other words, the game involves the design of objects.  We therefore
continue in the second semester with the use of programming video
games but supplement the course with a library for doing so in an
object-oriented style.  Figure~\ref{fig:world} gives an example
written in both the functional style and object-oriented style.

The key difference between these two programs is that the functional
program uses the {\tt 2htdp/universe} library, which provides a {\tt
  big-bang} form that consumes the initial state of the world and has
a declarative form of associating event-handler functions, while the
object-oriented program uses an alternative library developed for our
class: {\tt class/universe}.  It also provides a {\tt big-bang} form
but it consumes a single argument, the initial state of the world
represented as an object.  Event handlers are simply methods of this
object; for example, clock ticks trigger the {\tt on-tick} method.

The program on the left is the first program of the first semester,
while the one on the right is the first program of the second
semester.  Our approach is able to make the conceptual connection
between functional and object-oriented programming quite clear while
appealing to the familiar event-driven interactive programs developed
throughout the year.

The move to object-oriented style immediately and naturally leads to
designs that are enabled by organizing programs around objects.  For
example, the state pattern~\cite{samth:GOF} becomes useful almost
immediately.  The programs in figure~\ref{fig:world} animate a rocket
(rendered as a circle in this example) taking off.  An illustrative
follow-up exercise is to animate a rocket that \emph{lands}.  The
natural design is to have two variants for states of the rocket: one
for descending rockets and one for landed rockets (an example is given
in appendix~\ref{sec:appendix}).  While in the functional approach it
is easy to use the state-pattern for the \emph{data} representing a
rocket, it is more difficult to have states of \emph{behavior}.  The
typical solution adds conditionals to all of the event handlers.  In
the object-oriented approach, states of behavior are just as natural
as data.  It is therefore straightforward to design programs with easy
to observe invariants such as ``a landed rocket never changes
position.''  In the functional approach, even such simple properties
are more involved to establish, because all event handlers must be
inspected.

This approach also leads naturally to discussion of inheritance.
Often programs with multiple states wish to share the implementation
of some methods.  We first show that this can be accomplished at the
cost of minor boilerplate with delegation, and then show how
inheritance allows the programmer to avoid duplication and boilerplate
entirely.  Once inheritance is able to group identical methods,
overriding is a natural next step when some but not all of the
implementations are identical across the variants.

\begin{figure}
\begin{minipage}[t]{3.4in}
\begin{verbatim}
#lang bsl
(require 2htdp/image 2htdp/universe)

;; A World is a Number

;; on-tick : World -> World
(define (tick w)
  (add1 w))

;; to-draw : World -> Image
(define (draw w)
  (place-image 
    (circle 10 "solid" "red")
    w 200 (empty-scene 400 400)))

;; on-key : KeyEvent World -> World
(define (on-key k w) 10)

(big-bang 10
          [to-draw draw]
          [on-tick tick])
\end{verbatim}
\end{minipage}
\begin{minipage}[t]{3in}
\begin{verbatim}
#lang class/1
(require 2htdp/image class/universe)

;; A World is a (new world Number)
(define-class world
  (fields n)

  ;; on-tick : -> World
  (define (on-tick)
    (new world (add1 (this . n))))

  ;; to-draw : -> Image
  (define (to-draw) 
    (place-image 
     (circle 10 "solid" "red")
     (this . w) 200 (empty-scene 400 400)))

  ;; on-key : KeyEvent -> World
  (define (on-key k) (new world 10)))
  
(big-bang (new world 10))
\end{verbatim}
\end{minipage}
\caption{World programs}
\label{fig:world}
\end{figure}

\subsection{Language levels}

Our introduction to object-oriented programming is built on a series
of ``language levels'', each of which introduces additional features,
adding complexity to the programming model and expressiveness to the
programs.  Each language is  \texttt{class/}\textit{N}
for some \textit{N}, with features appearing in the following order.
\begin{enumerate}
  \setcounter{enumi}{-1}
\item Classes and Objects
\item Abbreviated notation for method calls
\item Super classes
\item Overriding
\item Constructors
\end{enumerate}

Several commonalities run through all of these languages.  First, they
are all purely functional; we do not introduce imperative I/O or
side-effects until after transitioning to Java in the second half of
the course.  Second, they all are a super set of the
\emph{Intermediate Student} language from \htdp, meaning that they
support higher-order functional programming and lists. 

One key principle that we adhere to in the design of the language
levels is that no features of the language are added purely to support
``software engineering'' concerns such as specification mechanisms.
Not only does that language not support declaring types or contracts,
but interfaces are described purely in comments.

This is not to say that interfaces and contracts are optional; in
fact, they are mandatory.  But the focus of the first part of the
course is on the fundamentals of object-orientation.  Teaching the use
of software engineering tools such as type systems, while vital, is a
topic which we defer to the second half of the course when we
transition to Java. 

We made this decision after experience in which students were confused
about the relationship between explicit interface specifications, type
systems, and the informal data definitions and contracts which
students are required to write for all methods.  After removing
interfaces from the language and making them purely a specification
construct, this confusion disappeared.

%% Fixme: add these sections back later

%\subsection{Universe}

%\subsection{Programming to interfaces}

\section{From principles to industrial languages}
\label{sec:industrial}

The transition from custom teaching languages to a professional
language takes place about half-way through the course.  At this
point, students already have experience with many of the essential
concepts of object-oriented programming. In particular: objects,
classes, fields and methods, dynamic dispatch, inheritance, and
overriding.

From this point, almost any language that students might encounter in
future co-op positions, summer internships, or upper-level courses
would be an appropriate follow-up.  Our course transitions to Java,
but C\#, Python, Ruby, Eiffel, or JavaScript would all work naturally.
The key lesson of the transition is that the fundamental principles
underlying object-oriented programming remain the same between
languages, and that learning a new language is primarily a matter of
mapping these concepts to specific constructs in the new language.  Of
course, particular languages also use unique specific mechanisms which
need to be taught to use the language effectively, but these are
rarely as vital as the cross-language principles.

We chose the half-way point as the time for transition based on
experience with earlier versions of this course. In particular, we
found that a later transition, while allowing us to present additional
concepts in a controlled environment, did not give students sufficient
time and experience with Java.  Subsequent classes found that students
were strong on fundamentals but weak on Java practice.  The other
alternative, transitioning earlier, would not provide sufficient time
to cover the fundamental topics before the transition.

\subsection{Functional Java}

The transition begins with replicating the object-oriented style of
our teaching languages in Java.  In particular, we do not introduce
mutation, for loops, or mutable data structures such as arrays or
\texttt{ArrayList}s until later in the semester.  Instead, students
design data representations using classes, with \texttt{interface}s
representing unions of data.  Additionally, we avoid mention of the
distinction between primitive and other values in Java, which is made
easier by not using standard libraries early.  An example of this
style of programming is presented in figure~\ref{fig:java}, repeating
the binary tree sum from the previous section.

\begin{figure}
\begin{verbatim}
import tester.*;

interface Tree {
    // sums the elements of this tree
    Integer sum();
}

class Leaf implements Tree {
    Integer v;
    Leaf(Integer v) { this.v = v; }
    public Integer sum() { return this.v; }
}

class Node implements Tree {
    Tree left; Integer v; Tree right;
    Node(Tree l, Integer v, Tree r) {
        this.left = l;
        this.v = v;
        this.right = r;
    }

    public Integer sum() {
        return this.left.sum() + this.v + this.right.sum();
    }
}

class Examples {
    void test_tree(Tester t) {
        t.checkExpect(new Leaf(7).sum(), 7);
        t.checkExpect(new Node(new Leaf(1),
                               5, 
                               new Node(new Leaf(0), 10, new Leaf(0))).sum(),
                      16);
    }
}
\end{verbatim}
\caption{Binary tree sum in the style of \emph{How to Design Classes}}
\label{fig:java}
\end{figure}

Comparing this figure to the previous example illustrates a number of the
differences that students are exposed to upon transition to Java.

\begin{enumerate}
\item Explicit representation of unions and interfaces in the
  language.  Previously, interfaces were simply described in stylized
  comments, following the \htdp approach.
\item Types are now specified as part of the program and are
  (statically) enforced.  Data definitions and interfaces can be
  transformed from the stylized comments into {\tt interface}
  definitions and method signatures annotated with types.  Students
  are taught the benefits of type systems, which impose syntactic
  restrictions sufficient to prove objects meet (the structural
  aspects of) their interface definitions.  Students also encounter
  the downside of types when they discover the type system cannot
  always follow valid reasoning about program invariants and may reject
  perfectly good programs.
\item Java syntax is substantially different and more verbose. For
  example, constructors must be defined explicitly.
\item The testing environment is somewhat different, and requires
  additional boilerplate, although we are
  able to use the JavaLib framework~\cite{local:java-world} to support testing with
  structural equality.
\end{enumerate}

\noindent
There are other differences which cannot be seen from a code snippet.

\begin{enumerate}
  \setcounter{enumi}{4}
\item Students must use a new development environment and compiler.
  In class, we primarily develop in a text editor and run the Java
  compiler at the command line.  In labs and on homeworks, students
  typically use the Eclipse IDE.
\item Installing and configuring libraries is now required.  Because
  we use a custom library for testing, students must cope with library
  installation and class paths on the first day.
\end{enumerate}

All but the first two of these changes are unrelated to the
fundamental lessons we hope to teach---the rest merely present
additional hurdles for students.  However, at this point in the
semester, the students are far better equipped to meet these
challenges.  They are already familiar with objects, classes, and the
other concepts we have covered.  They are also fully engaged in the
class, instead of making the transition in the midst of the transition
between semesters.  Finally, they have now been programming for 50\%
longer than they had at the start of the semester.

\subsection{Traditional Java}

Thanks to the preparation in the first half of the course, we can
cover OO programming in a functional subset of Java in a just a few
lectures.  We then increase the subset of the language we use to
encompass mutation, loops, and mutable data structures and introduce
the underlying design principles. We present \texttt{ArrayList}s,
followed briefly by arrays. Students use, and then implement, hash
tables as well as other mutable and immutable data structures.
Conventional input and output are treated only very briefly, as we
focus instead on both fundamentals and exercises making use of real
APIs such as hashing functions or Twitter posting.  Finally,
\texttt{while}, \texttt{for}, and for-each loops are presented,
following the methodology of \htdc which connects loops to stylized
use of recursive functions with accumulators, a technique the students
now have used for two semesters.

\subsection{Beyond Traditional Java}

Finally, at the end of the course, we are able to build on the two
major segments to examine less-well-explored topics in object-oriented
programming.  Typically, we cover the basics of representing objects
in a functional language, advanced OO techniques such as mixins and
prototypes, and a new OO language such as Ruby or JavaScript.
Additionally, we emphasize the ability to embed functional programming
in an OO context, using techniques such as the command pattern and the
visitor patterns.  Again, the key message is the transferability of
concepts across languages.

\section{Related work}
\label{sec:related-work}

Teaching programming principles in a functional style has a long
history, with Abelson and Sussman's \emph{Structure and Interpretation
  of Computer Programs}~\cite{dvanhorn:sicp} being a prominent
example.  Our work follows in the tradition of the \emph{Program by
  Design} (PbD)
project\footnote{\url{http://www.programbydesign.org/}} (previously
known as the \emph{TeachScheme!} project), which emphasizes a
systematic approach to program construction.

Since the introduction of functional-first curricula, and more
specifically in the Program by Design framework, numerous courses have
tackled the problem of transition.  Typically they, as we, transition
to Java in the second course.  We discuss first the approach developed
by some of the principal creators of PbD, and then other approaches.

\subsection{Program by Design and ProfessorJ}

The Program by Design project initially focused only on the first course,
with the second course typically taught in Java in
institution-specific ways. Subsequently, the pedagogical approach was
extended to Java, but without the tool support and textbook of the
first course.  An example of this approach is described by
\citet{dvanhorn:Bloch2000Scheme}, who presents the experience
integrating these courses at Adelphi.  He reports that ``many of
Java's concepts could be introduced more easily in a second course
than a first.''

With these lessons in mind, the PbD project set out to apply
the lessons of teaching languages and IDE support to Java, as well as
to present the approach to object-oriented programming in textbook
form. ProfessorJ~\cite{dvanhorn:Gray2003ProfessorJ} is the resulting
system, accompanying the draft textbook \emph{How to Design
  Classes}~\cite{local:htdc}.  In parallel to our course, Northeastern
teaches the remainder of its computer science majors following this
approach.

ProfessorJ and \emph{How to Design Classes} maintain many of the ideas
of the first course.  In particular, ProfessorJ brings language levels
to Java, in an attempt to smooth the transition for students from the
first course and provide more helpful feedback.  ProfessorJ is also
embedded in the DrRacket IDE, increasing familiarity for the students
and supporting tools such as an interactive read-eval-print loop.

However, the ``day 1'' transition from the student languages used with
\htdp to \profj is too abrupt and too large.
Most significantly, changing languages from the first semester
immediately rather than simply adding a new concept confuses too many
issues for students.  On the first day of a \htdc-based course,
students see object-orientation, a new programming paradigm; Java, a
new language with new syntax, and a static type system, a crucial but
orthogonal concept.
In contrast, our course presents just one of these concepts on the
first day, but  covers all of them by the end of the semester.

\profj also takes on the dual challenges of implementing Java as
well as subsetting it.  This ultimately resulted in both a limited
Java environment as well as the eventual abandoning of the tool since
it was too difficult to maintain, let alone keep up with advances in
Java.  

Committing to Java on the first day, regardless of the environment
provided to students, has significant limitations.  First, the
syntactic and semantic heaviness of Java is a burden for beginning
students, and discourages interactive and experimental programming.
The very first chapter of \htdc discusses the fixed size of Java
integers, a topic avoided entirely in the first course.  Second, by
committing to a particular industrial-strength language, it closes off
possibilities in the curriculum.  Third, it commits entirely to the
new paradigm, making it more difficult for students to compare the
approaches.

Since \profj is no longer available, students are faced with an even
starker change on the first day.  Even with a student-oriented
environment such as DrJava or BlueJ~\cite{dvanhorn:Allen2002DrJava,
  dvanhorn:Hsia2005Taming}, students must learn an entirely new tool,
along with new libraries.
If the course uses a typical professional development environment such
as Eclipse, students must also contend with compilation, loss of
interactivity, and subtle issues such as classpaths, none of which are
fundamental to the concepts that the course focuses on.

\subsection{Other transitions}

Not every curriculum that begins with \htdp transitions to Java after
the first course.  \citet{dvanhorn:Ragde2008Chilling} describes a
second course that includes both more advanced work in Scheme beyond
teaching-oriented languages as well as low-level programming in C,
taught to computer science majors at University of Waterloo.  Radge's
course intentionally does not use student-oriented languages, although
the recently-developed C0 language~\cite{local:c0} could provide such
a language. Other discussions of functional programming in the first
year~\cite{dvanhorn:Chakravarty2004Risks} do not discuss the problems
of transition.

\subsection{Other approaches to Java}

The problems of teaching Java in introductory courses have been
well-explored; we mention only a few related directions here.
DrJava~\cite{dvanhorn:Allen2002DrJava} and
BlueJ~\cite{dvanhorn:Hsia2005Taming,dvanhorn:Kolling2003} are
introductory environments for Java, which alleviate some but not all
of the drawbacks we have outlined.
For example, both of these systems immediately present students with
(1) type systems and (2) Java syntax, and (3) do not support the image
values and exact numeric values that we rely on in our course.

Several teaching-oriented graphics libraries for Java have been
proposed~\cite{dvanhorn:Bruce2001Library,dvanhorn:Alphonce2003Using},
but these are significantly more complex than the graphics and
interaction libraries we are able to use in the introductory language
we present.

\section{Experience and outlook}
\label{sec:conclusion}

We have now completed the third iteration of this course, teaching
approximately 35 students each time.  Our experience has been
uniformly positive, and the students have gone on to significant
success in the subsequent courses, despite the curriculum differing
from what the bulk of Northeastern University computer science majors
take.  Anecdotally, the class has also had notable success in the
recruitment and retention of female students, as compared to the other
versions of the second-semester course. However, the classes are
sufficiently different as to make a precise comparison impossible.

The course has provided a vantage point to introduce topics that will
be taken up later in the curriculum.  We present types, contracts,
invariants, and properties of functions, all of which tie into both
the concurrent course on logic and computation, as well as later
classes on formal methods.  The emphasis on
representation-independence and interfaces both tie into later classes
on software engineering, as well as preparing students for algorithms
and data structures courses.  Finally, the use of interactive and
distributed systems connects to later courses on operating systems and
networks.

Despite our success, much remains to be done.  Type systems are a
fundamental concept, but their introduction accompanies the rest of
Java.  Developing a typed version of our introductory languages would
allow a smoother introduction of this idea.  

Our class's use of Eclipse could also be improved by first
transitioning to a pedagogically-oriented Java environment, but we
have not evaluated the specific options.  Alternatively, introducing
Java-like syntax for the teaching languages we have developed would
help tease apart the difficult transitions still present in the course.

Finally, the Java portion of the class does not continue the use of
``World''-style interactive graphical programming, although a version
of the ``World'' library has been developed for Java
\cite{local:java-world}.  Instead, our course focuses on coverage of
standard Java libraries, as well as introductory algorithmic and data
structure topics.  Continuing to use World-style programming in
motivating examples might be valuable for continuity between the two
halves of the course.

\subsection*{Acknowledgments}

Matthias Felleisen's approach to pedagogy and passion for
undergraduate teaching has inspired this work from the beginning.
CCIS Dean Larry Finkelstein entrusted two postdocs with the redesign
of a key undergraduate course, which made this experiment possible.
Our teaching assistants, Dan Brown, Asumu Takikawa, and Nicholas
Labich, as well as the tutors and graders, contributed enormously to
the success of our courses. Finally, and most importantly, our
students at Northeastern for the last three years have put up with a
curriculum in progress, and the opportunity to teach them has been
truly rewarding.

% \nocite{*}
\bibliographystyle{eptcs}
\bibliography{dvh-bibliography,local,sth-bibliography}

\appendix
\section{Worlds and the State pattern}
\label{sec:appendix}

\begin{alltt}
#lang class/1
(require 2htdp/image class/universe)

;; A World is one of 
;; - (new landed-world)
;; - (new downworld Number)
(define-class landed-world

  ;; to-draw : -> Image
  (define (to-draw) 
    (place-image 
     (circle 10 "solid" "red")
     390 200 (empty-scene 400 400))))

(define-class downworld
  (fields n)

  ;; on-tick : -> World
  (define (on-tick)
    (cond [(zero? (this . n))
           (new landed-world)]
          [else
           (new world (sub1 (this . n)))]))

  ;; to-draw : -> Image
  (define (to-draw) 
    (place-image 
     (circle 10 "solid" "red")
     (this . w) 200 (empty-scene 400 400)))

  ;; on-key : KeyEvent -> World
  (define (on-key k) (new world 400)))
  
(big-bang (new downworld 400))
\end{alltt}

\end{document}